\documentclass{jps-cp}
\usepackage{txfonts} 
\usepackage{ulem}
\usepackage{color}

\title{Pressure-Temperature Phase Diagram of $\alpha$-Mn}

\author{Takaaki \textsc{Sato}$^{1}$, Kazuto \textsc{Akiba}$^{1}$, Shingo \textsc{Araki}$^{1}$, and Tatsuo C. \textsc{Kobayashi}$^{1*}$}

\inst{$^{1}$ Department of Physics, Okayama University, Okayama 700-8530, Japan}

\email{kobayashi@science.okayama-u.ac.jp}

\recdate{September 17, 2019}

\abst{
Electrical resistivity and ac-susceptibility measurements under high pressure were carried out in high-quality single crystals of $\alpha$-Mn. 
The pressure-temperature phase diagram consists of an antiferromagnetic ordered phase ($0<P<1.4$ GPa, $T<T_{\rm N}$), a pressure-induced ordered phase ($1.4<P<4.2$-4.4 GPa, $T<T_{\rm A}$), and a paramagnetic phase. 
A significant increase was observed in the temperature dependence of ac-susceptibility at $T_{\rm A}$, indicating that the pressure-induced ordered phase has a spontaneous magnetic moment. 
Ferrimagnetic order and parasitic ferromagnetism are proposed as candidates for a possible magnetic structure.
At the critical pressure, where the pressure-induced ordered phase disappears, the temperature dependence of the resistivity below 10 K is proportional to $T^{5/3}$.
This non-Fermi liquid behavior suggests the presence of pronounced magnetic fluctuation.
}

\kword{$\alpha$-Mn, electrical resistivity, ac-susceptibility, quantum critical point}

\begin{document}
\maketitle

\section{Introduction}
Most elemental transition metals generally form simple crystals such as body centered cubic (bcc), hexagonal close-packed (hcp), and face centered cubic (fcc) crystals.
However, $\alpha$-Mn forms a significantly complex bcc structure that has 58 atoms in the cubic unit cell ($I\bar{4}3m$, A12 structure) with four crystallographically non-equivalent Mn sites.
This structure is identical to the $\chi$ phase found in multi-component intermetallic compounds.
In the case of $\alpha$-Mn, the electron configurations in the $s$, $p$, and $d$ orbitals are different for each of the Mn sites.
Thus, one can assume that Mn atoms at different sites behave as if they have different atomic sizes;
this is could be a possible reason why $\alpha$-Mn has such a complex crystal structure \cite{Brewer}.
The $\alpha$ phase is the most stable phase at room temperature; at higher temperatures, there are sequential structural transitions to the $\beta$ (1000-1368 K: simple cubic), $\gamma$ (1368-1406 K: fcc), and $\delta$ (1406-1517 K: bcc) phases.
Under high pressure, one can expect a denser and more symmetrical crystal structure than in the $\alpha$ phase.
Surprisingly, however, the $\alpha$ phase is known to be stable up to 165 GPa \cite{Fujihisa}.

$\alpha$-Mn is known to exhibit an antiferromagnetic (AFM) transition at $T_{\rm N}=95$ K.
A neutron diffraction study on single crystalline $\alpha$-Mn showed the magnetic moment on each Mn site, \textit{i.e.}, 2.05 $\mu_{\rm B}$ for site I, 1.79 $\mu_{\rm B}$ for site II, $\sim$0.6 $\mu_{\rm B}$ for site III, and $\sim$0.3 $\mu_{\rm B}$ for site IV,
and the noncolinear AFM spin structure \cite{Yamada}.
Here, $\mu_{\rm B}$ represents the Bohr magneton.
Correspondingly, the previous nuclear magnetic resonance study conducted in a zero-magnetic field reported the site-dependent resonance frequencies for each Mn site \cite{Yamagata}.

In the resistivity measurements under high pressure, the emergence of a pressure-induced ordered phase is reported above 1.5 GPa.
As a result, the pressure-temperature ($P$-$T$) phase diagram shows a characteristic two-step structure \cite{Takeda, Miyake}.
At ambient pressure, the resistivity increases just below $T_{\rm N}$, which reflects a change of the band structure through the AFM transition.
However, a transition to the pressure-induced ordered phase does not accompany the increase of resistivity at the transition temperature $T_{\rm A}$.
At a pressure of approximately 5 GPa, $T_{\rm A}$ vanishes, and the proportional coefficient of $T^2$ shows divergent behavior.
This suggests that a quantum critical point (QCP) exists at $\sim 5.0$ GPa.
However, the details of the magnetic structure in the pressure-induced ordered phase and the magnetic fluctuation expected near the QCP remain unclear at present. 

In order to discuss the QCP at $\sim$5.0 GPa, clarifying the magnetic order in the high pressure phase is of prime importance. 
In the present study, the transport and magnetic properties of single crystalline $\alpha$-Mn under high pressure are investigated by means of resistivity and ac-susceptibility measurements. 
The emergence of a huge increase in ac-susceptibility at $T_{\rm A}$ is observed for the first time, which indicates that the pressure-induced ordered phase accompanies the spontaneous magnetic moment.
Near the critical pressure, at which the pressure-induced ordered phase vanishes, the resistivity shows $T^{5/3}$ dependence below 10 K.
This non-Fermi liquid behavior can be regarded as a hallmark of magnetic fluctuation.

\section{Experiment}
Single crystalline $\alpha$-Mn crystals were synthesized by the Pb-flux method, which was recently reported in \cite{Fukuhara}.
Mn (99.999\%) and Pb (99.9999\%) with a molar ratio of 2:98 were placed in an alumina crucible and sealed in a quartz ampoule under the partial pressure of argon.
After the crystals were initially heated to $800\ {}^\circ\mathrm{C}$, the melt was cooled to $320\ {}^\circ\mathrm{C}$ for over 300 h.
Then, the flux was removed by a centrifuge separator.
The obtained crystals were polyhedral with a maximum dimension of 700 $\mu$m, as shown in the inset of Fig. \ref{f1}.
These polyhedral crystals were confirmed to be single crystals by means of X-ray diffraction analysis.
They were then shaped into thin plates by mechanical polishing for resistivity measurements. The typical residual resistivity ratio (RRR) of the investigated samples was approximately 16,
which was defined by the resistivity ratio between 300 K and 2 K.
Indenter-type \cite{Kobayashi} and opposite anvil-type \cite{Kitagawa} pressure cells were used to generate hydrostatic pressure.
Daphne 7474 oil was used as a pressure medium.
Pressure in the sample space was determined based on the superconducting transition temperature of Pb placed together with the samples.
Resistivity measurements were performed by the standard four-probe method.
Electrical contact was made using silver paste.
The ac-susceptibility measurements were performed by using coaxial primary and secondary coils (0.5 mm diameter and 1 mm long) placed in the sample space of a pressure cell.
The volumes of the sample and the Pb used in the ac-susceptibility measurements were $1.1\times 10^{-4}$ cm$^{-3}$ and $2.4\times 10^{-5}$ cm$^{-3}$, respectively.
The absolute value of ac-susceptibility was calibrated assuming that all volume fractions of Pb showed the Meissner effect.
As the sample space was limited, we performed the measurements without a cancellation coil.
Measurements down to 50 mK were performed using a dilution refrigerator.

\section{Results}

\begin{figure}[]
\centering
\includegraphics{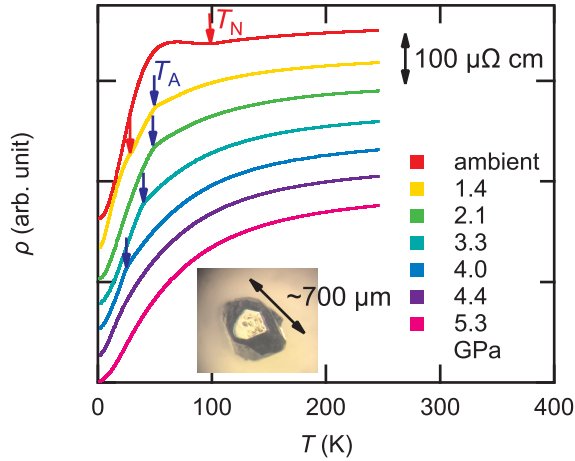}
\caption{Temperature dependence of the resistivity $\rho$ at representative pressures with transition temperatures $T_{\rm N}$ (red arrows) and $T_{\rm A}$ (blue arrows).
Each curve is vertically shifted for clarity.
The inset shows a picture of single crystalline $\alpha$-Mn synthesized by Pb-flux method.}
\label{f1}
\end{figure}

The temperature dependence of the electrical resistivity ($\rho$) of $\alpha$-Mn under high pressure is shown in Fig. \ref{f1}. 
Each curve is vertically shifted for clarity.
As reported previously \cite{Takeda, Miyake}, the N\'{e}el temperature is confirmed to be $T_{\rm N}$ = 95 K at ambient pressure. 
The resistivity increases just below $T_{\rm N}$ then decreases as temperature decreases, which results in a hump in the $\rho$-$T$ curve. 
The increase in the resistivity just below $T_{\rm N}$ shows the decrease of carrier density due to the partial disappearance of the Fermi surface. 
The application of pressure lowers $T_{\rm N}$, and
causes another phase transition above 1.4 GPa, which has been ascribed to $T_{\rm A}$ in the previous reports \cite{Takeda, Miyake}. 
The resistivity decreases just below $T_{\rm A}$ with decreasing temperature, which causes a kink in the $\rho$-$T$ curve. 
In the data, at 1.4 GPa, both a hump at $T_{\rm N}$ and a kink at $T_{\rm A}$ are observed simultaneously.
This indicates that successive transition from pressure-induced ordered phase to AFM phase takes place at 1.4 GPa.
$T_{\rm A}$ decreases with further application of pressure and can be traced up to 4.1 GPa.

\begin{figure}[t]
\centering
\includegraphics{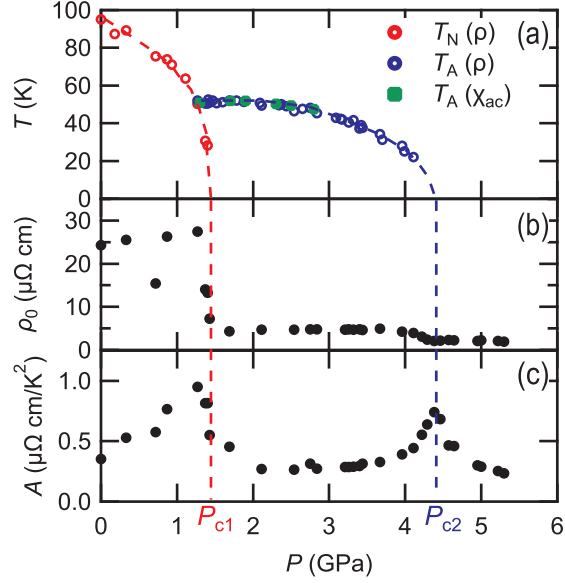}
\caption{(a) $P$-$T$ phase diagram of $\alpha$-Mn.
Red and blue symbols indicate $T_{\rm N}$ and $T_{\rm A}$ determined by resistivity measurements,
and green symbols indicate $T_{\rm A}$ determined by ac-susceptibility.
(b) Pressure dependence of residual resistivity $\rho_0$. 
(c) Pressure dependence of coefficient $A$ of $T^2$ term of the resistivity.}
\label{f2}
\end{figure}

The $P$-$T$ phase diagram is obtained from the above experimental results, as shown in Fig. \ref{f2} (a).
Here, $T_{\rm N}$ and $T_{\rm A}$ are defined as the midpoint of the jump in the temperature dependence of $d\rho/dT$. 
The AFM phase disappears at 1.4 GPa, while the pressure-induced ordered phase appears. 
The observation of the successive transition at 1.4 GPa firstly establishes the phase boundary between the AFM phase and the pressure-induced ordered phase. 
Above 1.4 GPa, $T_{\rm N}$ vanishes with slight increases of pressures less than 0.5 GPa, indicating that the phase boundary is nearly perpendicular to the horizontal axis.
The critical pressure at zero temperature is determined as $P_{\rm c1}$ = 1.4 GPa.
The $T_{\rm A}$ is not sensitive to the pressure just above $P_{\rm c1}$, and decreases with further application of pressure.
Although $T_{\rm A}$ can be determined up to 4.1 GPa, the kink in the $\rho$-$T$ curve cannot be traced above 4.1 GPa.

The pressure dependence of residual resistivity $\rho_0$ is shown in Fig. \ref{f2} (b).
The $\rho_0$ shows two steps at two critical pressures ($P_{\rm c1}$ and $P_{\rm c2}$ in Fig. \ref{f2}), which is consistent with previous results \cite{Takeda, Miyake}. 
The difference of the band structure among the three ground states causes the steps in the pressure dependence
of $\rho_0$.
As can be seen in Fig. \ref{f2} (b), the pressure-induced ordered phase seemingly disappears at $P_{\rm c2}\sim~$ 4.2-4.4 GPa, which is lower than that of the previous report \cite{Takeda}.

\begin{figure}[t]
\centering
\includegraphics{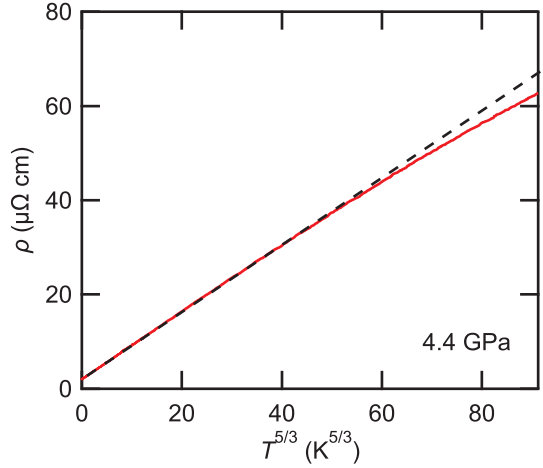}
\caption{Temperature dependence of resistivity at 4.4 GPa plotted against $T^{5/3}$.
The temperature range shown is from 0 K to 15 K.
Linear broken line is a guide for the eye.}
\label{f3}
\end{figure}

\begin{figure}[t]
\centering
\includegraphics{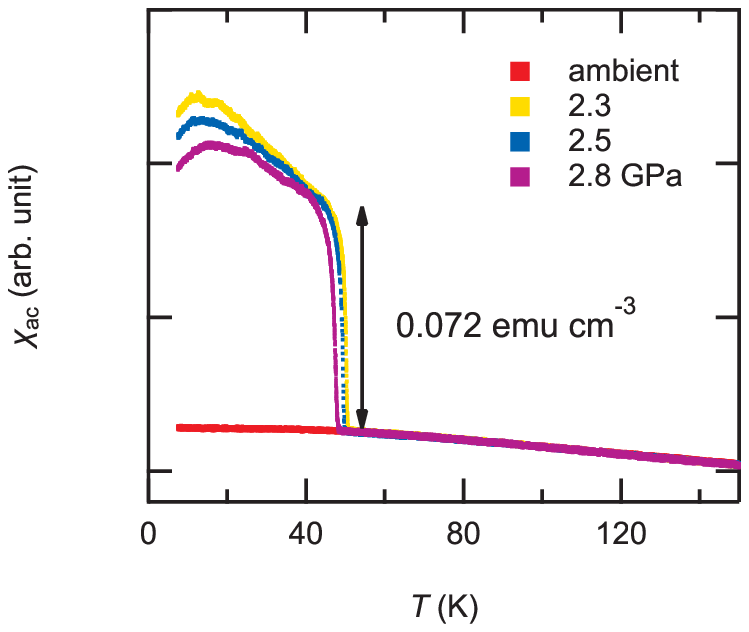}
\caption{Temperature dependence of ac-susceptibility $\chi_{\rm ac}$ at ambient pressure, 2.3, 2.5, and 2.8 GPa.
The vertical scale is estimated from the Meissner effect of Pb inserted together with the sample.}
\label{f4}
\end{figure}

The pressure dependence of the coefficient $A$ of the $T^2$ term of the resistivity (referred to as $A$-value, hereafter) in the lowest temperature region is shown in Fig. \ref{f2} (c).
The $A$-value peaks near $P_{\rm c1}$ and $P_{\rm c2}$.
The $A$-values are determined in the low temperature limit.
The temperature dependence of resistivity is proportional to $T^{\sim 1.9}$ at 1.4 GPa (not shown),
which is close to the Fermi liquid-like behavior.
At 4.2 GPa, on the other hand, temperature dependence shows $T^{5/3}$ dependence in the wide temperature range from 50 mK to 10 K. 
This non-Fermi liquid behavior indicates notable magnetic fluctuation in this pressure region.

The temperature dependence of the ac-susceptibility ($\chi_{\rm ac}$) of $\alpha$-Mn under high pressure is shown in Fig. \ref{f4}.
Although an anomaly in resistivity has been known to exist at $T_{\rm N}$ = 95 K at ambient pressure \cite{Takeda, Miyake}, 
it is not observed in the ac-susceptibility with the sensitivity in the present measurement. 
In contrast, a significant increase in the susceptibility is observed at around 50 K under high pressure (above 2 GPa), indicating that spontaneous magnetization occurs in the pressure-induced ordered phase.
As plotted in Fig. \ref{f2}(a), the transition temperatures determined from the rising of $\chi_{\rm ac}$ correspond well with those from resistivity.
As the ac-susceptibility is the reversible part of the initial susceptibility in the magnetization vs. magnetic field curve, we cannot estimate the magnitude of the ordered moment at the present stage.

\section{Discussion}
The absence of the pressure-induced structural phase transition was reported in $\alpha$-Mn up to 165 GPa \cite{Fujihisa}. 
Thus, the crystal structure of $\alpha$-Mn consists of four sites
(referred to as site I, II, III, and IV in the introduction part) even under high pressure. 
As it is difficult to assume a sign change of the exchange interaction at 1.4 GPa,
the ferromagnetic order can be excluded as a candidate for the pressure-induced ordered state.
A previous calculation showed the volume dependence of the magnitude of the magnetic moments at each site \cite{Hobbs}. According to the calculation, the moments at site IV disappear below 12.0 \AA$^3$/atom, whereas the moments at other sites survive down to 9-10 \AA$^3$/atom, corresponding to 60-140 GPa. 
It is possible that the disappearance of the AFM order at $P_{\rm c1}$ is related to the disappearance of the moment at site IV.
The calculation also suggested that the spin structure changes from noncollinear AFM to collinear AFM owing to the disappearance of the moments at site IV. 
However, spontaneous magnetization is expected to be absent in an ideal collinear AFM order.
Thus, this scenario cannot explain the present experimental result.
The ferrimagnetic order formed by the different magnetic moments at four sites or the parasitic ferromagnetism in the antiferromagnetic order can be suggested as the candidate of the origin of the spontaneous moment observed in the present study. 

Near the QCP where the magnetic fluctuation develops, non-Fermi liquid behavior was observed, in which the resistivity deviates from $T^2$ dependence. 
The self-consistently renormalized spin fluctuation (SCR) theory predicts that the temperature dependence of resistivity at QCP is proportional to $T^{3/2}$ in the AFM case and $T^{5/3}$ in the ferromagnetic case \cite{Moriya}. 
As can be seen in Fig. \ref{f3}, the temperature dependence of the resistivity at 4.4 GPa does not obey $T^{2}$; instead, it shows $T^{5/3}$ dependence, which indicates pronounced magnetic fluctuation in the vicinity of $P_{\rm c2}$.
This behavior corresponds well with the temperature dependence at the ferromagnetic QCP expected in SCR theory.
The origin of this ferromagnetic power law is an open question at the present stage.

\section{Conclusion}
Electrical resistivity and ac-susceptibility measurements were carried out on single crystalline $\alpha$-Mn crystals under high pressure.
The pressure-temperature phase diagram consists of the antiferromagnetic ordered phase, pressure-induced ordered phase, and paramagnetic phase.
The precise boundaries of these phases are clarified.
In the ac-susceptibility measurements, the emergence of an significant increase at the transition temperature to the pressure-induced ordered phase is observed for the first time, indicating the existence of spontaneous magnetic moment.
Ferrimagnetic order and parasitic ferromagnetism are proposed as candidates for a possible magnetic structure.
At the critical pressure, where the pressure-induced ordered phase disappears, the temperature dependence of resistivity shows the non-Fermi liquid behavior of $T^{5/3}$
over a wide temperature range of 50 mK to 10 K, which is due to the pronounced magnetic fluctuation.

\section{Acknowledgement}
The authors thank Dr. Hiromi Ota at Division of Instrumental Analysis, Okayama University  for the X-ray single crystal structural analyses.
This work was supported in part by Grants-in-Aid for Scientific Research (No. 18K03517 and No. 18H04323) provided by the Ministry of Education, Culture, Sports, Science and Technology (MEXT) of Japan.

\end{document}